\documentclass[a4paper,11pt]{article}
\usepackage{graphicx}  
\usepackage{epsfig}

\newcommand{\z}[1]{\ensuremath{\mathbf{#1}}}

\graphicspath{{figures_ACTA/}}

\begin{document}
\begin{center}
\ \\
{\large Active control of scattered acoustic radiation:\\a real-time implementation
for a three-dimensional object}\\
\ \\
short title:  {Real-time control of scattered noise}\\
\ \\
Emmanuel~Friot, R\'egine~Guillermin, Muriel~Winninger\\

{CNRS - Laboratoire de M\'ecanique et d'Acoustique,\\
$31$ chemin Joseph Aiguier, $13402$ Marseille, France}\\
\end{center}

\ \\
\textbf{Abstract}

This paper presents an active noise control experiment
designed to validate a real-time control strategy for reduction
of the noise scattered from a three-dimensional body.
The control algorithm relies on estimating the scattered noise by linear filtering
of the total noise measured around the body; suitable filters are identified
from off-line measurements. A modified Filtered-Error Least-Mean-Squares
algorithm then leads to the adaptive filters which drive the secondary sources.
The paper provides the numerical simulations using a Boundary Element Method which helped
in designing a feasible experiment in an anechoic chamber with a limited number of control sources.
Eventually a real-time pure-tone implementation with 14 ordinary loudspeakers
and a large body is shown to yield on average a 10~dB reduction of the scattered noise at the error sensors, 
which is close to the optimum reduction predicted by the numerical simulations
for the sensor arrangement.\\

 
\section{Introduction}

In theory, active control can attenuate the noise scattered from a reflecting body
(see \cite{Nelson}). 
In practice however, very few experiments regarding active control of scattered noise have been reported
because no generic control strategy nor specific technology are available.
Non-invasive Huyghens' sources, which theoretically lead to perfect control, are difficult
to build using real sources; control with one source was successfully
implemented in the case of waves coming from one direction \cite{Mangiante}, but
arranging a full layer of Huyghens' sources is not easily feasible and a control
strategy valid for any kind of sources is desirable.
Global control of the body surface vibration is another idea to reduce the scattered
noise, but this requires a three-dimensional implementation of new technologies
(see for example \cite{Howarth,Nayrolles}) 
and so far has not allowed in practice control in the shadow zone of a scattering body.
Local active control of the impedance can largely increase the sound absorption
at the surface of a scattering body (see \emph{e.g}
\cite{Olson,Orduna,Furstoss,Scandrett}), but local absorption
does not suppress the forward scattering, which is a global effect depending on the shape
as much as on the surface properties of a scattering body.

A control strategy has been proposed recently to control the scattered noise using ordinary
noise sources and sensors such as, in air, loudspeakers and microphones \cite{Friot}.
This control strategy does not require information about
the propagation medium, and two-dimensional (2D) numerical simulations as well as a preliminary 1D
experiment in a duct have shown that it could yield reduction of the scattered noise
without {\it a priori} information about the incident wave.
The present paper describes the design and implementation of
a 3D experiment of scattered noise reduction using this control strategy.
Additions to the previous theory and
numerical simulations are used to elaborate a feasible experiment with a
limited number of channels. In the end, the paper provides results of real-time
control with 14 secondary sources of the noise scattered in an anechoic chamber
by a rigid parallelepiped. 

Active control of scattered noise is of interest in protecting submarines
or underwater mines from sonar detection \cite{Howarth}, but it seems that the
required number of actuators prevents a practical application to these cases in the
near future. Cancellation of the reflections from the walls of anechoic chambers
at low frequency is another potential application (see \cite{Friot-CFM}).
Even if the latter is in the authors' line of sight, the control experiment presented
in this paper does not pretend to be a scale-model of any of these applications;
the experiment was designed to validate the real-time implementation of the control strategy
indroduced in \cite{Friot}.

In section 2, the basic strategy for active control of the scattered noise is
recalled, which includes the corrections and the additions to the previous theory
which were necessary for a successful 3D experiment.  
The control strategy relies on an estimation of the scattered
noise by linear filtering of acoustic pressure measurements around the reflecting body,
which is suggested by the the Green's integral representation of the scattered noise.
The filters which map the total noise to the scattered noise do not depend on
the incident noise. They can be identified
from {\it in situ} measurements of the Frequency Response Functions from the
control sources to microphones around the body. In this way, the Green's function for
the propagation medium and the dynamics of the secondary sources are automatically integrated
into the design of the optimal controller. The estimated scattered noise at locations
around the reflecting body is then used as a set of error signals for the Filtered-Error
Least-Mean-Square adaptation (FELMS, see \cite{Elliott}) of the filters which drive the control
sources in real-time.

In section 3, numerical simulations with a Boundary Element Method code 
are used to compute the optimal reduction of the scattered noise as a function
of the number of sensors and actuators.
2D simulations have shown that 3 microphones per wavelength
on the reflecting body surface were needed for a large attenuation of the scattered noise
going to all directions \cite{Friot}. In the paper, this is is confirmed
by 3D numerical simulations in the case of a rigid $1.56\times0.94\times0.5\;\textrm{m}^3$
parallelepiped surrounded by actuators and sensors. 
Thus, plenty of sensors are
required for a large body at high frequency, whereas off-the-shelf digital control
systems are unable to handle more than a few dozen channels. Therefore, the design
of a feasible experiment in an anechoic chamber is a compromise solution between
the required number of channels, the size of the reflecting body and the frequency.
In the paper, further simulations show that a limited arrangement involving 14 loudspeakers
and 18 microphones on the same level can lead to some reduction of the scattered noise
at a working frequency close to 300~Hz. 
The limited number of channels does not allow the control of the scattered noise throughout the
whole 3D space, but it is sufficient for control in more than one direction.
These simulations form the basis of the design of a real-time control experiment. 

In section 4, the experimental set-up and results are presented. The experiment was arranged
in the LMA large anechoic chamber and real-time computations were performed on the multi-processor
control system COMPARS. The scattered noise with and without control was measured
using 16 microphones around the reflecting parallelepiped. It was not used for the real-time
computation of the control source inputs. Ultimately a 10~dB reduction of the scattered
noise at the error sensors was achieved on average, which is close to the optimal control
for the given arrangement of sources according to the numerical simulations in section~3.

\section{Theory of the control algorithm}

Besides secondary sources, a typical set-up for Active Noise Control includes
error sensors spanning the minimization area and reference sensors detecting
the incoming noise in advance. The input signals for the secondary sources can be
computed by filtering the reference signals, and steepest-descent adaptive algorithms 
often lead to nearly optimal control filters without
requiring an accurate model of the plant \cite{Elliott}.
In the case of scattered noise, this feedforward scheme cannot be implemented
in a straightforward manner because no error sensors can measure the scattered noise only.
The theory of the control algorithm given here is intended to allow real-time computation
of the scattered noise from measurements using ordinary sensors.
Fig.~\ref{photo} displays the kind of set-up considered in this section:
a reflecting body is surrounded by inner noise sensors and by secondary sources.
A set of outer sensors is also used for preliminary measurements but not for real-time
computation of the source inputs.

\subsection{Mapping of the total noise to the scattered noise}

The acoustic pressure inside a volume $V$ bounded by a surface $S$
can be written, at angular frequency $\omega$, as (see \textit{e.g.} \cite{Morse}):
\begin{equation} \label{eq:Green0}
\!\!p(\z{r},\omega) = 
\int\!\!\!\!\!\int\!\!\!\!\!\int_V 
fG(\z{r},\z{r}_0,\omega)dV_0 +
 \int\!\!\!\!\!\int_S \left[   
G(\z{r},\z{r}_0,\omega) \frac{\partial}{\partial n_0}p(\z{r}_0,\omega)
- p(\z{r}_0,\omega)\frac{\partial}{\partial n_0}G(\z{r},\z{r}_0,\omega) \right] dS_0
\end{equation}
where $f$ is the noise source distribution in $V$, $G$ a Green's function and
$\frac{\partial}{\partial n_0}$ denotes the component of the normal gradient
to the surface and pointing away from $V$. Volume $V$ may be infinite, in this case
the integral over the part of surface $S$ surrounding $V$ at infinity 
vanishes because of the properties of Green's functions.
In the special case when $S$ encloses a scattering body
and $G$ is the Green's function for the propagation medium \emph{without} the body,
the surface integral which appears in Eq.~(\ref{eq:Green0}) vanishes when the body is removed
because the total pressure is given only by the volume integral. Therefore,
the scattered pressure $p_s$, accounting for the body presence, is:
\begin{equation} \label{eq:Green}
p_s(\z{r},\omega) = \int\!\!\!\!\!\int_S \left[
G(\z{r},\z{r}_0,\omega) \frac{\partial}{\partial n_0}p(\z{r}_0,\omega)
- p(\z{r}_0,\omega)\frac{\partial}{\partial n_0}G(\z{r},\z{r}_0,\omega) \right] dS_0
\end{equation}
Because the contribution of the incident field to the surface integral
is zero, the scattered pressure $p_s$ could be substituted
to $p$ in the right hand side of this equation. However, using $p$
in equation (\ref{eq:Green}) shows that a linear operator maps the total acoustic
pressure all around the scattering body to the scattered pressure at any place
beyond surface $S$. This operator does not depend on the incident field.
Once it has been computed of identified from experimental data,
the scattered pressure can be computed, whatever the incident waves are, from
measurements of the total acoustic pressure which can be performed with ordinary microphones.

In classical textbooks on Active Noise Control (\emph{e.g.} in \cite{Nelson}), 
Eq.~(\ref{eq:Green0}) is meant to show that acoustic radiation can be cancelled
by laying out Huyghens' sources at the scattering surface.
In the present paper, Eq~(\ref{eq:Green0}) is only used to show that minimization
signals accounting for the scattered radiation can be computed by linear filtering of
the total radiation, without making any assumption about the secondary sources.
However, if a set of arbitrary sources would lead to perfect cancellation of scattered radiation,
the secondary field would then be equivalent to the one produced by Huyghens' sources
at the scattering surface.

Discretization of the integral in Eq.~(\ref{eq:Green}) leads to:
\begin{equation} \label{eq:Green2}
p_s(\z{r},\omega) \approx \sum_{k=1}^{N} \left[ g_k(\z{r},\omega)p(\z{r}_k,\omega)
+ h_k(\z{r},\omega)\frac{\partial}{\partial n_k}p(\z{r}_k,\omega)\right]
\end{equation}
This means that, for active control, error signals accounting for the scattered noise
can be computed by linear filtering of total noise measurements
with a finite set of ordinary noise sensors.
If filters $g_k$ and $h_k$ are known, no information whatsoever is required regarding
the primary noise sources.

If $S$ is the surface of a rigid body, the pressure gradient term vanishes in
(\ref{eq:Green2}) and only pressure microphones are required for practical computations.
In the case of a non-rigid body, \cite{Friot} stated that gradient microphones
were necessary. In fact this can be alleviated if the body has a locally reacting surface
$S$ with admittance $\beta(\z{r},\omega)$. In this case Eq.~(\ref{eq:Green2}) is written as:
\begin{equation} \label{eq:Green3}
p_s(\z{r},\omega) \approx \sum_{k=1}^{N} \left[ g_k(\z{r},\omega)-\rho j\omega\beta(\z{r}_k,\omega)
h_k(\z{r},\omega)\right]p(\z{r}_k,\omega) = \sum_{k=1}^{N} \hat{g}_k(\z{r},\omega) p(\z{r}_k,\omega)
\end{equation}
where $\rho$ is the medium density,
and again only pressure microphones are required for estimating the scattered noise.
Gradient microphones or microphone pairs are only required for bodies
with non-locally reacting surfaces, and so will no longer be considered in the paper
so as to alleviate the derivation of the real-time control algorithm. However, with
practical applications in mind, it must be noticed that vibrating surfaces are
usually not locally reacting surfaces.

Equation (\ref{eq:Green3}) gives access to the scattered noise at any location around the body.
\cite{Friot} stated that, if active control cancels the scattered pressure at
a sufficient number of sensors meshing a minimization surface around the body,
then it reduces it everywhere outside. In fact this is only true when the propagation
medium delimited by the sensors is infinite. In the case of minimization for a finite
medium (e.g. for a measurement area inside an anechoic room where loudspeakers
are used to cancel the wall reflections), zero pressure at the minimization boundary does not imply
zero pressure inside the surface at the resonance frequencies of the volume demarcated
by the minimization sensors: the pressure may not be zero at the resonance frequencies
of a finite volume with Dirichlet boundary condition.
In this case, pressure gradient microphones are also required
as minimization sensors. However it
has been reported in \cite{Ise} that breaking the symmetry in the position of the pressure
sensors can also lead to the cancellation of the pressure gradient on the surface,
which ensures that the noise is cancelled outside the surface.

\subsection{Identification of the scattering filters}

If the Green's function for the propagation medium is known,
discretization of Eq.~(\ref{eq:Green}) gives the linear filters $\hat{g}_k$
in Eq.~(\ref{eq:Green3}), but a better idea, valid for the general case,
is to identify the filters from noise data collected before control.
To this effect, a set of incident waves can be generated by the control sources
and a set of microphones can be used to measure the scattered noise by 
subtracting the noise measured without the body to that measured
with the body. Inversion of Eq.~(\ref{eq:Green3}) evaluated at the set of
microphones then leads to coefficients $\hat{g}_k$. For details see \cite{Friot},
where it was stated that the scattering filters
were uniquely determined provided the number of secondary sources was larger
than the number of microphones on the surface enclosing the reflecting body.
Although this is algebraically true, the data collected using the set-up
of section 4 led to an ill-conditioned inversion of Eq.~(\ref{eq:Green3}).
A regularization technique had to be implemented, which as a side-effect
relaxed the constraints on the relative number of actuators and sensors.
Whatever these numbers were, regularization was also justified by the fact that
the discrete set of scattering filters $\hat{g}_k$ arose from the discretization
of a continuous linear operator in Eq.~(\ref{eq:Green}).
The filters should not have been very sensitive to the specific data collected
for identification, which was enforced by regularization.
In practice, the control results given in section 4 were obtained
after the Tikhonov regularization of the inverse problem leading to
the scattering filters (see \cite{Tikhonov}). 
When no regularization was imposed, the estimation of the scattered noise
was too poor for control, which was probably due to insufficient measurement data
for inverse identification of the scattering filters.

It must be noted that this identification of the scattering filters from off-line
inversion of experimental data only suits to situations with stationary environment.
In the case of a changing environment, periodic identification or update of the
filters would be required, which would be much more difficult to implement in practice.

\subsection{Real-time implementation of the control algorithm}

The vector $\z{e}$ of error signals that real-time control has to minimize is the sum,
at the minimization locations and at discrete time $n$, of the \emph{scattered} primary noise
vector $\z{p}_s^{(1)}$ and of the \emph{total} secondary noise $\z{p}^{(2)}$.
The secondary noise generated by the control sources is also the sum of a scattered noise
$\z{p}_s^{(2)}$ and an incident noise $\z{p}_i^{(2)}$:
\begin{equation} 
\z{e} (n) = \z{p}_s^{(1)}(n) + \z{p}^{(2)}(n) = \z{p}_s^{(1)}(n) + \z{p}_s^{(2)}(n) + \z{p}_i^{(2)}(n)
\end{equation}
Let \z{G} be a matrix of Finite Impulse Responses (FIR) which approximates, in the time domain,
the scattering filters $\hat{g}_k$ which map the vector $\z{p}$ of the total pressure at the
inner sensors in to the scattered pressure
$\z{p}_s$ at the minimization sensors. Hence:
\begin{equation} 
\z{p}_s(n) = \z{p}_s^{(1)}(n) + \z{p}_s^{(2)}(n) = \z{G}*\z{p}(n)
\end{equation}
where $*$ denotes the time convolution product.
Let $\z{H}_i$ be the FIR approximation of the Frequency Response Functions (FRFs) from the vector of secondary source
inputs \z{u} to the vector of incident secondary noise: $\z{p}_i^{(2)}(n)= \z{H}_i*\z{u}$.
Matrix $\z{H}_i$ can be identified before control. The vector of error signals is eventually
expressed as a real-time function of sensor measurements \z{p} and known outputs \z{u}:
\begin{equation} \label{eq:errors}
\z{e}(n) = \z{G}*\z{p}(n) + \z{H}_i*\z{u}(n)
\end{equation}

Real-time control using a FELMS algorithm also requires a vector \z{x} of reference signals 
which detect the primary noise before it reaches the minimization locations.
Vector of control signals \z{u} is then computed by filtering reference vector \z{x} by a matrix of
adaptive FIR filters \z{W}, namely $\z{u}(n) = \z{W}*\z{x}(n)$. If \z{H} denotes a FIR approximation
of the matrix of secondary paths from $\z{u}$ to $\z{e}$,
the tap coefficients $\{\z{W}_0, \ldots, \z{W}_k, \ldots, \z{W}_{K-1}\}$ of filters \z{W}
can be updated according to (FELMS algorithm, {\it cf.} \cite{Elliott}):
\begin{equation} \label{eq:FELMS}
\z{W}_k(n+1) = \z{W}_k(n) - \alpha  \left[ \z{H}^{\dag}*\z{e(n-\tau+1-k)}\right] \z{x}(n)
\end{equation}
where $\z{H}^{\dag}$ is the transposed matrix of the \emph{time-reversed} responses in \z{H},
delay $\tau$ is the length of FIR \z{H}, and $\alpha$ is a convergence coefficient.
Inserting Eq.~(\ref{eq:errors}) into Eq.~(\ref{eq:FELMS}) finally gives
the adaptation formula for control of the scattered noise:
\begin{equation} \label{eq:fin}
\z{W}_k(n+1) = \z{W}_k(n) - \alpha  \left( \z{H}^{\dag}*
	\left[\z{H_i}*\z{u}(n-\tau+1-k) + \z{G}*\z{p}(n-\tau+1-k)\right] \right) \z{x}(n)
\end{equation}
\cite{Friot} gave alternative expressions for the error signals and for
the updating formula. Equation~(\ref{eq:fin}) was used to implement the experiments
in section~4 because it appeared to be the least sensitive to measurement errors.
The convolution products between stationary filters in (\ref{eq:fin}) can be performed off-line,
and in the end the real-time computations require only twice as much memory and time as a usual FELMS due
to the one additional real-time convolution.

\section{Numerical simulations}

This section presents 3D numerical simulations which evaluate the
performances of the control strategy of the scattered noise. The simulations were
performed with two objectives in mind: the first one was the confirmation
in a 3D case of the "rule of thumb", derived from 2D simulations in \cite{Friot},
which stated that 3 sensors per wavelength are required for estimation and control
of the scattered field. The second objective was the numerical simulation
of various arrangements, with a limited number of sensors, in order to determine a compromise
configuration for an experimental study with off-the-shelf control units.
The scattering body chosen for the
simulations given here was a $1.56\times 0.94\times 0.5\;\textrm{m}^3$ rigid parallelepiped. 
This choice corresponded with the object eventually used in the experimental study
presented in section~4.

\subsection{Computation of the scattered noise}

The scattering of an acoustic wave by a three-dimensional object can be
computed using a Boundary Element Method (see {\it e.g.} \cite{Kirkup}).
In this study the object was assumed to be rigid, a Neumann boundary condition was imposed
on its surface. The surface was meshed with at least seven triangular elements
per wavelength. Computations were performed at several frequencies,
each one associated to a specific mesh.
For example the object surface was meshed with 864 triangular elements for 280~Hz
and 2,656 elements for 500~Hz.

\begin{figure}[!ht]   
\begin{center}  
\includegraphics[width=14cm]{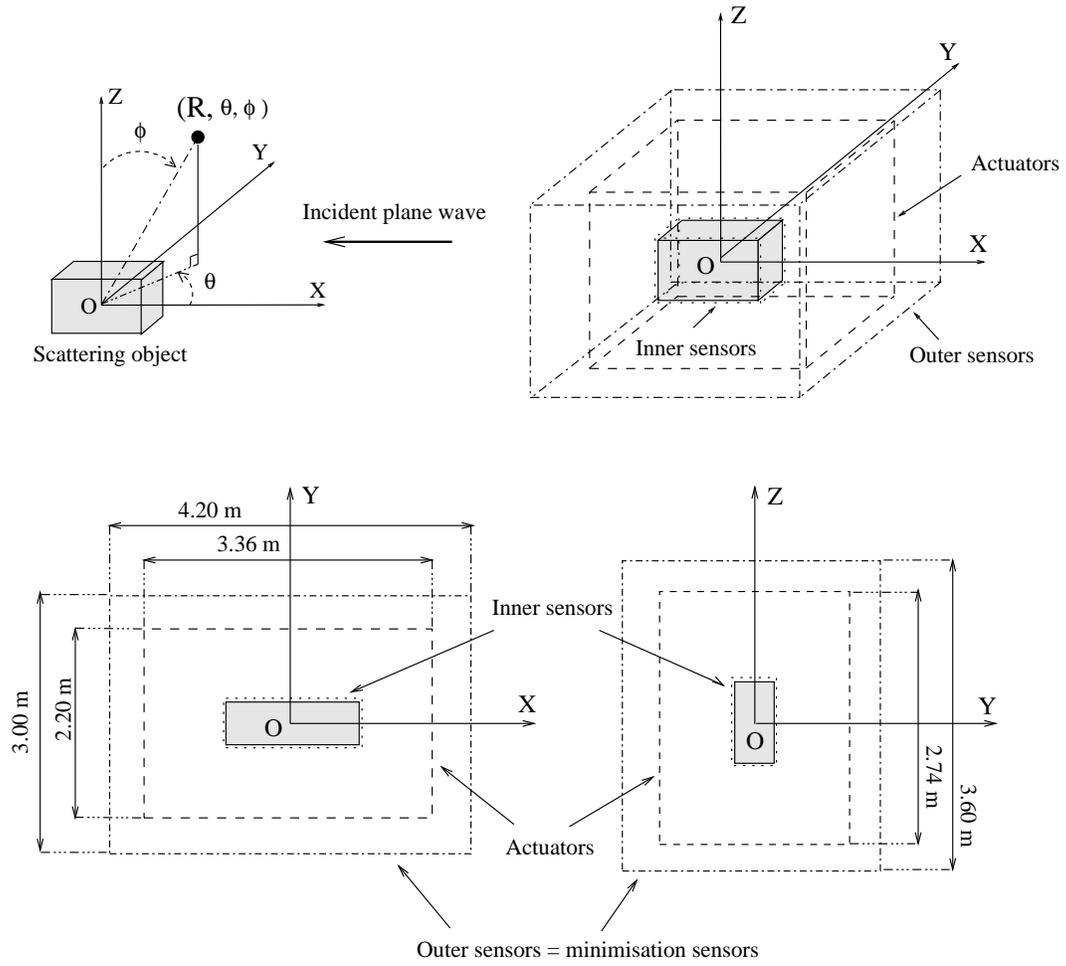} 
\vspace*{2cm} \caption{Configuration of sensors and actuators for numerical tests}  
\label{configsimu}  
\end{center}
\end{figure}

In the numerical study, the secondary sources for active control were considered as
monopole sources. The incident noise coming from the primary source
was a monochromatic plane wave with peak level 1~Pa (\textit{i.e.} with a 0.71~Pa
Root-Mean-Square value, which amounts to a 91~dB Sound Pressure Level)
coming from the right, with incidence angle $\theta=0$
(see Fig.~\ref{configsimu}). In order to estimate the control efficiency,
the scattered field with and without control was computed on a sphere of radius $R$
with its center at $O$, enclosing the scattering object. 
A far as the sensor and actuator positions are concerned, numerous combinations are possible
for numerical simulations.
Only a few of these, which should give an idea of how the control fares and
how to design a real-time experiment, are considered below.

\subsection{Simulation of full 3D control}

In \cite{Friot}, a numerical study in a 2D case showed that the strategy outlined
in section~2 was requiring minimum numbers of inner and outer sensors.
For an accurate estimation of the scattered
noise at the outer minimization sensors, more than 2 inner sensors per wavelength
were required near the surface of the scattering body.
For reduction of the scattered radiation in all directions, more than 2
outer error sensors per wavelength were required on the minimization surface
surrounding the scattering body. In general, using as many actuators as sensors
allows perfect noise cancellation at the error sensors because perfect control
involves the inversion of the transfer matrix from the actuators to the error sensors.
However in practice the number and the position of the actuators must be
selected in order to maximize the condition number of the secondary matrix, and
setting up more sensors than actuators is usually a good idea.
It was also shown in \cite{Friot} than 3 sensors per wavelength
on both measurement surfaces were enough for a reduction of the scattered radiation which was
larger than 20~dB in all directions. 
These numbers, which link up with theorems
on spatial sampling and aliasing, are common place in studies on global active noise control
(see \cite{Nelson}). The meshes of sensors may appear coarse when comparing
with Boundary Element Method computations, where six elements per wavelength are usually required.
However low-order BEM assume a crude behaviour of the acoustic pressure along an element,
which requires meshes with more nodes than for field noise sampling. 
In this section, a few simulation results are given to
confirm that, in the 3D case of scattering by a parallelepiped body, about three
sensors per wavelength allow efficient control of the scattered radiation in all directions.

For numerical simulations of full 3D control, the sensors and actuators were positioned
onto three parallepipedic surfaces as displayed in Fig.~\ref{configsimu}.
So as to ensure 3 sensors and actuators per wavelength at frequencies around 280~Hz, which
was the frequency where experiments were intended,
the distance between two inner (or outer) microphones had to be less than
$\lambda/3 \simeq 40 cm$. This meant that 54 inner and 324 outer microphones
were needed for a regular meshing of the parallepipedic surfaces.
This last number could have been reduced by bringing the outer microphones closer to the object,
but the real size of the secondary sources imposed a minimum distance in the later real-time experiments,
which was taken into account in the numerical simulations.
Simulations showed that a mesh of 302 actuators was adequate for control at the 324 outer sensors,
control with 54 inner sensors, 324 outer sensors and 302 actuators will be referred to
as "configuration (1)" below.

\begin{figure}[!ht]  
\begin{center}  
\includegraphics[width=14cm]{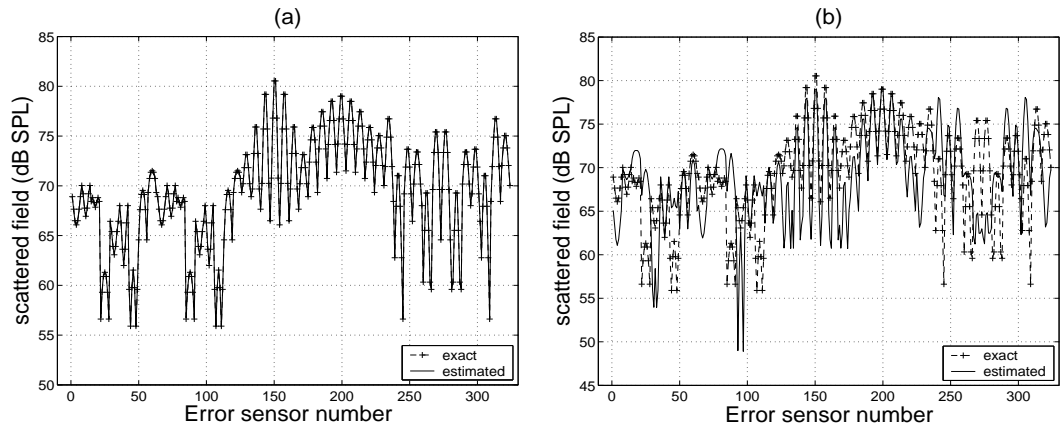} 
\vspace*{2cm} \caption{Scattered field estimation at the outer sensors,
(a) for 54 inner sensors, (b) for 28 inner sensors} 
\label{estimation3}  
\end{center}  
\end{figure}

Firstly, Fig.~\ref{estimation3} displays the scattered field at 280~Hz at the outer error sensors,
as well as its estimation
from measurement of the total pressure at 54 or at 28 inner sensors. For this figure
the scattering estimation filters were identified as explained in section 2.
The figure shows that 54 inner sensors at
the parallelepiped surface allow an accurate estimation of the scattered noise,
whereas the estimation is poor when only 28 sensors are used. Therefore, in this 3D case,
more or less than three inner sensors per wavelength are required for an accurate discretization
of equation (\ref{eq:Green}). This was confirmed in other numerical simulations where
the working frequency was increased while the number of sensors remained unchanged. 

\begin{figure}[!ht]  
\begin{center}  
\includegraphics[width=14cm]{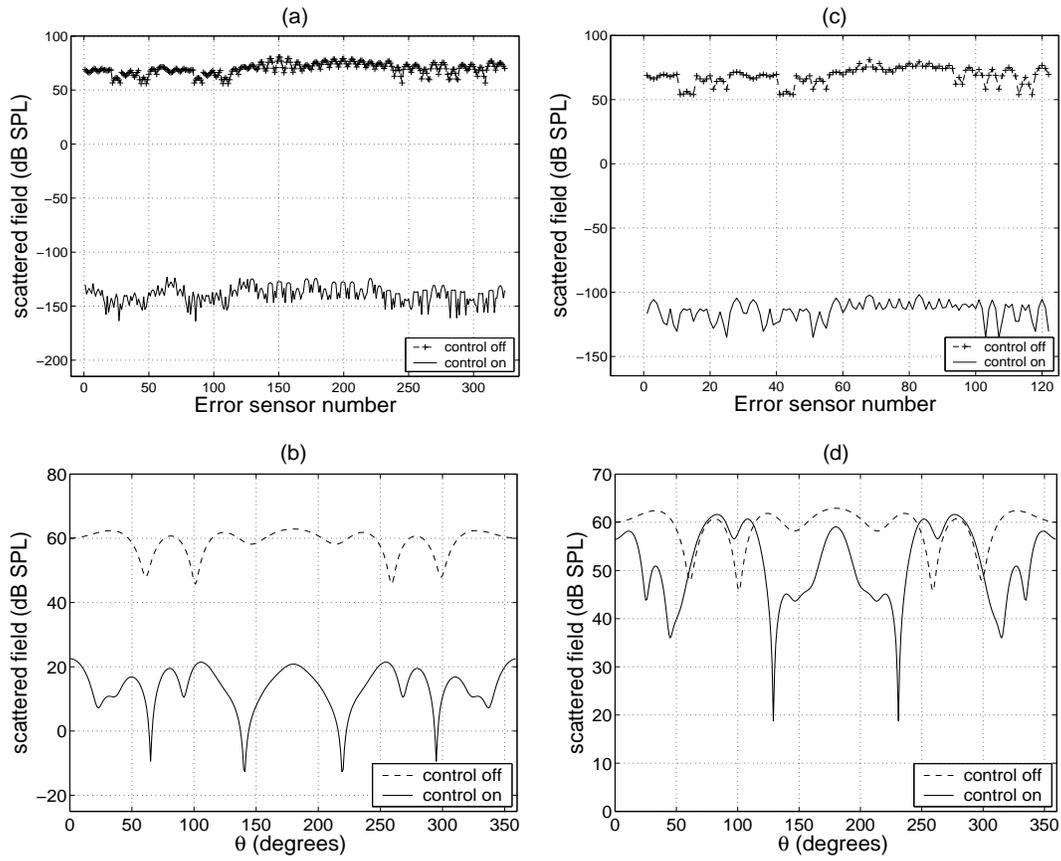} 
\vspace*{2cm} \caption{
Far-field effect of control of the exact scattered noise at the outer sensors:
(a) scattered noise at a set of 324 error sensors,
(b) far-field scattered noise in the plane $z=0$ with 324 outers sensors,
(c) scattered noise at a set of 122 error sensors,
(d) far-field scattered noise in the plane $z=0$ with 122 outers sensors}
\label{nombremicvirt}  
\end{center}  
\end{figure}

Secondly, Fig.~\ref{nombremicvirt} displays the scattered field for control at 324 
error sensors (with 302 sources) or at 122 error sensors (with 136 sources
---~and regularization of the secondary matrix inversion). The figure shows
the scattered noise at the outer error sensors as well as in the far field of the scattering body
on a circle in the plane $z=0$ and at distance $R=10\lambda$. Control does cancel
the scattered noise at the two sets of error sensors, but it leads to far-field reduction in all directions
only in the case with 324 error sensors. With 122 error sensors, the scattered noise
passes through the mesh of error sensors. Again, this shows that 3 error sensors per wavelength
on the minimization surface are required for global noise control outside the surface. This was
also confirmed by numerical simulations at increasing frequencies.

\begin{figure}[!ht]  
\begin{center}  
\includegraphics[width=14cm]{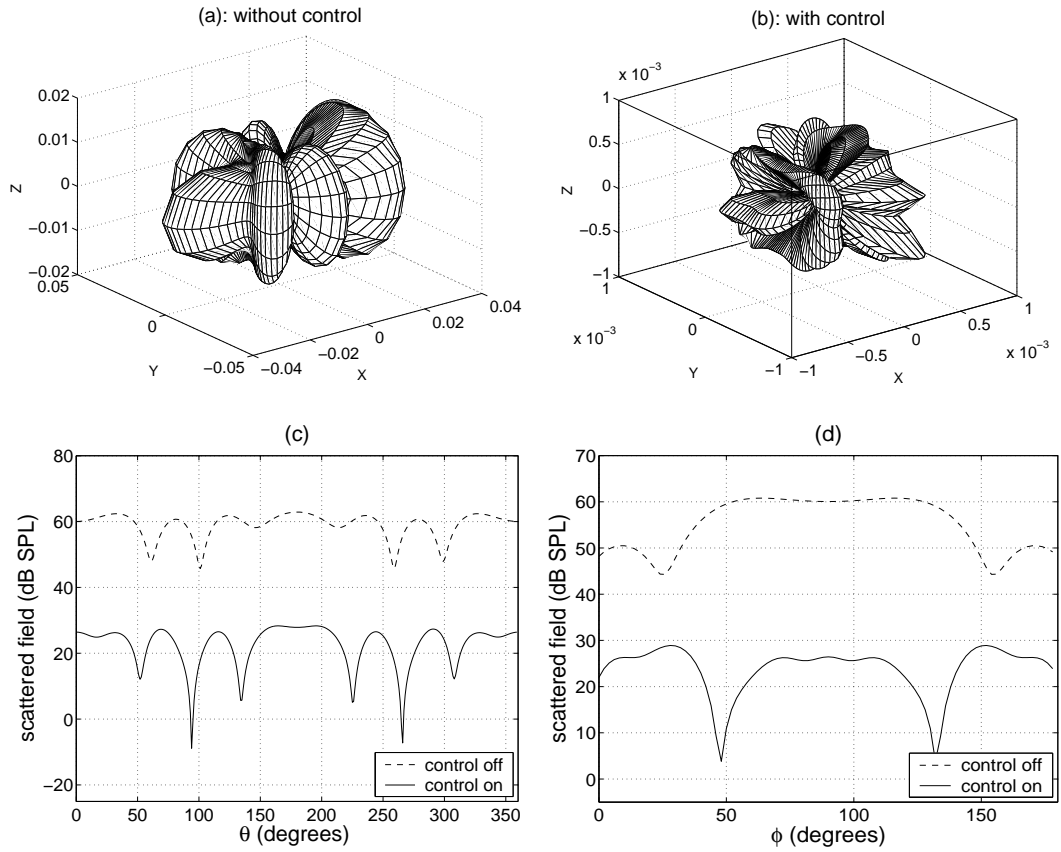} 
\vspace*{2cm} \caption{Configuration (1) - scattered noise in far-field
(a) without control,
(b) with control of the estimated scattered noise at the outer sensors,
(c) in the plane $z=0$,
(d) in the plane $y=0$}  
\label{2803d}  
\end{center}  
\end{figure}

Lastly, Fig.~\ref{2803d} shows various plots of the scattered field,
with and without optimal control in configuration (1), computed on a sphere with radius $R=10 \lambda$.
Fig.~\ref{2803d}(a) is a 3D plot of the scattered pressure without control
and \ref{2803d}(b) with control. The scales in Fig.~\ref{2803d}(a) and
Fig.~\ref{2803d}(b) are dramatically different, with
the scattered noise being reduced by more than 25~dB in all directions.
Fig.~\ref{2803d}(c) and Fig.~\ref{2803d}(d) display the control results
on two perpendicular planes as a function of the azimut and elevation angles. 
An efficient reduction of the scattered noise has been achieved in the
far-field, which confirm the rule, establised in 2D, of three sensors per wavelength on the inner and outer
surface for successful implementation of the control strategy of section 2.
The performances of estimation and control are summarized in Table~1, where
index $\left\| \sum_{k} g_k p^{(1)}(\z{r}_k) - \z{p}_{s}^{(1)} \right\|$
accounts for the error in the estimation of the scattered noise, index $\|e\|$
accounts for the for residual error at the error sensor.

\subsection{Simulation of a limited 3D set-up}

Multipurpose real-time units currently allow the control of noise at a few dozen error
sensors. 
Numerical simulations were performed in order to design a 3D experiment with reduction
of the scattered noise in several directions in spite of a limited number of channels.

\begin{figure}[!ht]   
\begin{center}  
\includegraphics[width=13cm]{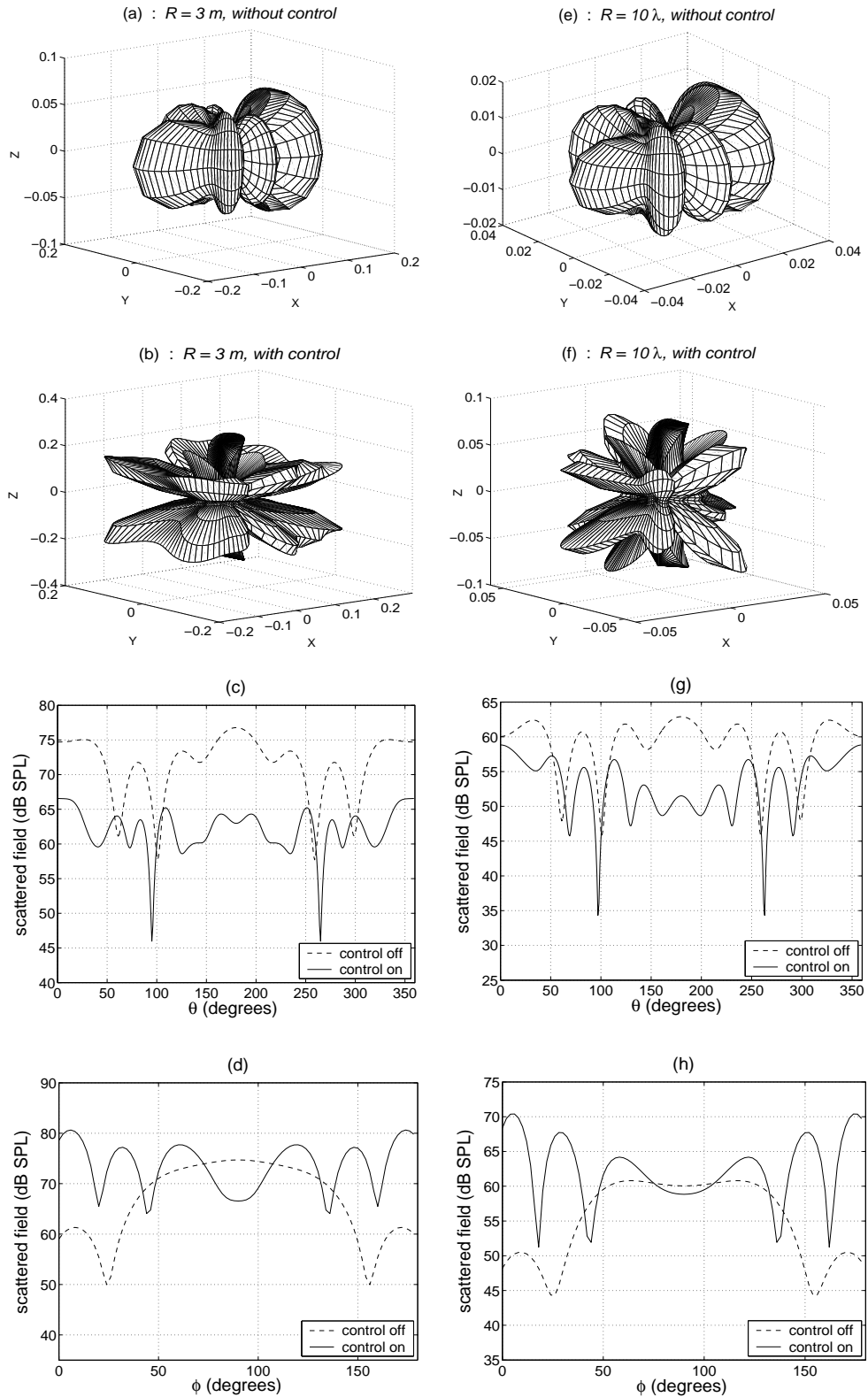} 
\caption{Configuration (2) - scattered noise:
(a) near the error sensors without control,
(b) near the error sensors with control,
(c) in the plane $z=0$ at $R=3\;m$,
(d) in the plane $y=0$ at $R=3\;m$,
(e) in the far-field without control,
(f) in the far-field with control,
(g) in the far-field in the plane $z=0$,
(h) in the far-field in the plane $y=0$}  
\label{280deg1}  
\end{center}  
\end{figure}

A first possible limited set-up was designed by considering actuators and sensors
in the single plane $z=0$. Simulations were performed with 58 outer sensors and
56 actuators.
For this case named "configuration (2)", the scattered field was computed with and without control near the
outer microphones at $R=3$~m and in far field at $R=10 \lambda$, and the results
are shown in Fig.~\ref{280deg1}. It appeared that control was efficient on
the outer microphones (Fig.~\ref{280deg1}(b) and Fig.~\ref{280deg1}(c)) but, as expected, the scattered 
field was not attenuated in other directions around the object; it was even amplified 
in some other directions, see Fig.~\ref{280deg1}(d). However, although
control might have been strictly localized in a zone near the outer
microphones, computations at $R=10\lambda$ showed that the control was still
effective in far-field beyond the outer microphones. The scattered noise was 
still attenuated by 5 to 10~dB in the plane where $z=0$, see Fig.~\ref{280deg1}(e) to~\ref{280deg1}(h).
It must be noted that the scattered field is reduced both in the shadow and in the bright zone
of the body. 

\begin{figure}[!ht]   
\begin{center}  
\includegraphics[width=14cm]{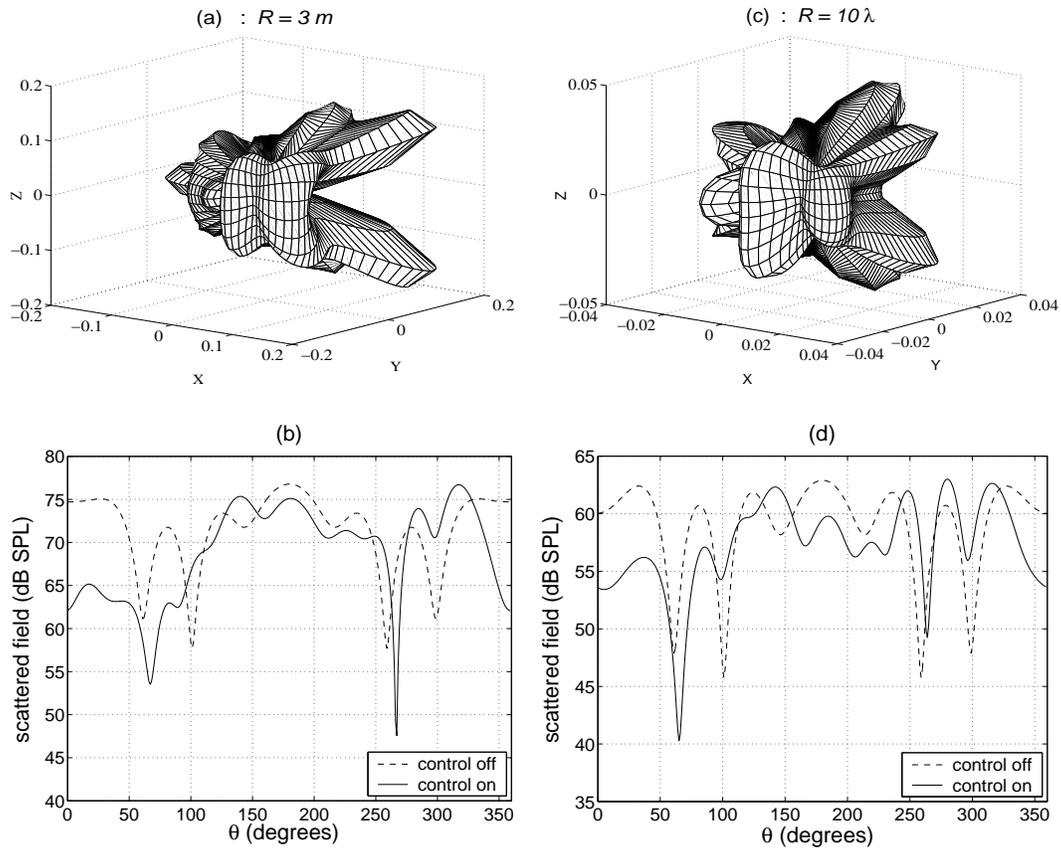} 
\vspace*{2cm} \caption{Configuration (3) - scattered noise:
(a) near the error sensors with control,
(b) in the plane $z=0$ at radius $R=3\;m$,
(c) in the far-field with control,
(d) in the far-field in the plane $z=0$}  
\label{280deg2}  
\end{center}  
\end{figure}

\begin{figure}[!ht]  
\begin{center} 
\includegraphics[width=14cm]{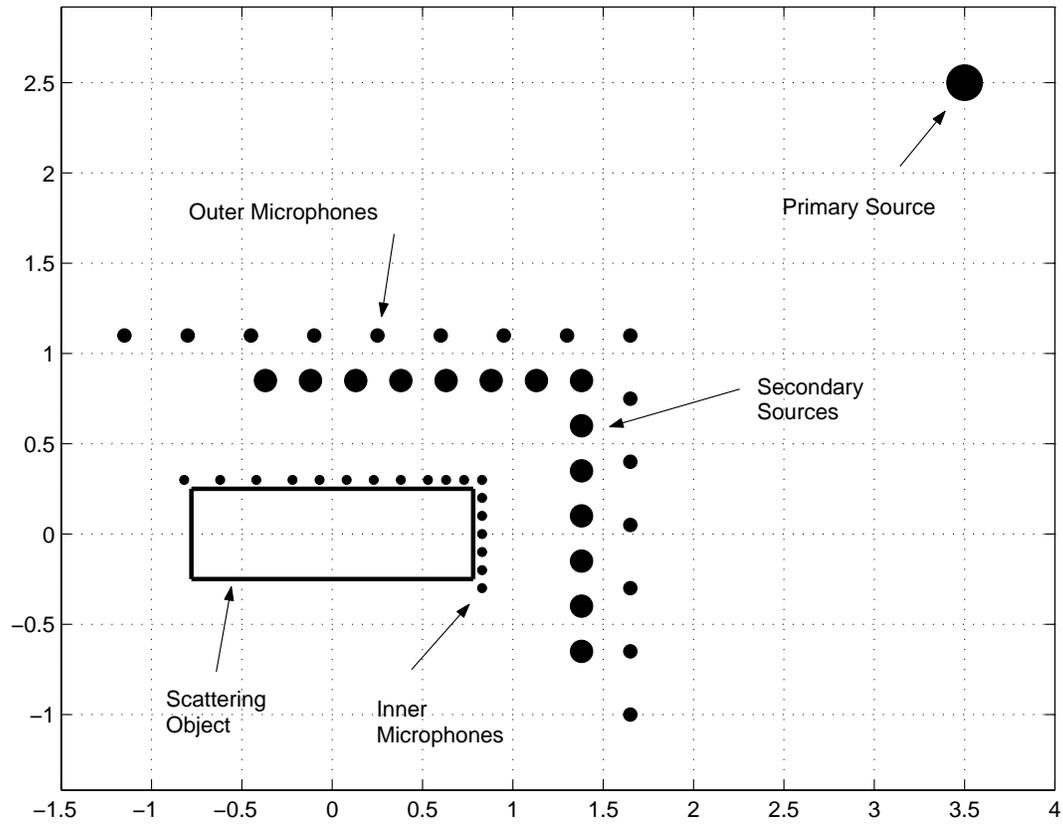}    
\vspace*{2cm} \caption{Arrangement of actuators and sensors in the experimental set-up}  
\label{schemamanip}  
\end{center}  
\end{figure}

The LMA system COMPARS and the loudspeakers available for real-time experiments
are limited to the processing of 14 actuators, 18 inner sensors and 16 outer error sensors.
Furthermore, the identification of the scattering filters suggested in section~2.2
supposes that it is possible to remove or insert the scattering object
without changing the sensor or actuator positions. With a heavy and bulky
object this is more easily done in practice if the object is not completely surrounded by microphones.
This led to the experimental set-up depicted in Fig.~\ref{schemamanip} where
the sensors and the actuators were again in the plane $z=0$ but within a limited angular domain
in the bright zone of the scattering body.
This arrangement named "configuration (3)" was eventually selected for the real-time experiment
presented in section~4 and pictured in Fig.~\ref{photo}.
Fig.~\ref{280deg2} displays the results of the numerical simulation for
this last case, where a 10~dB attenuation was achieved on average in the control zone.
Finally it must be noticed
that the SPL of the scattered noise, which was approximately 65~dB at the outer microphones,
was much less than the SPL of the incident noise. This means that the scattered noise can
be difficult to measure or to estimate accurately in practice,
which was confirmed by the subsequent experiments.

\section{Experimental study}

\subsection{Experimental set-up}

\begin{figure}[!ht]   
\begin{center}  
\includegraphics[width=14cm]{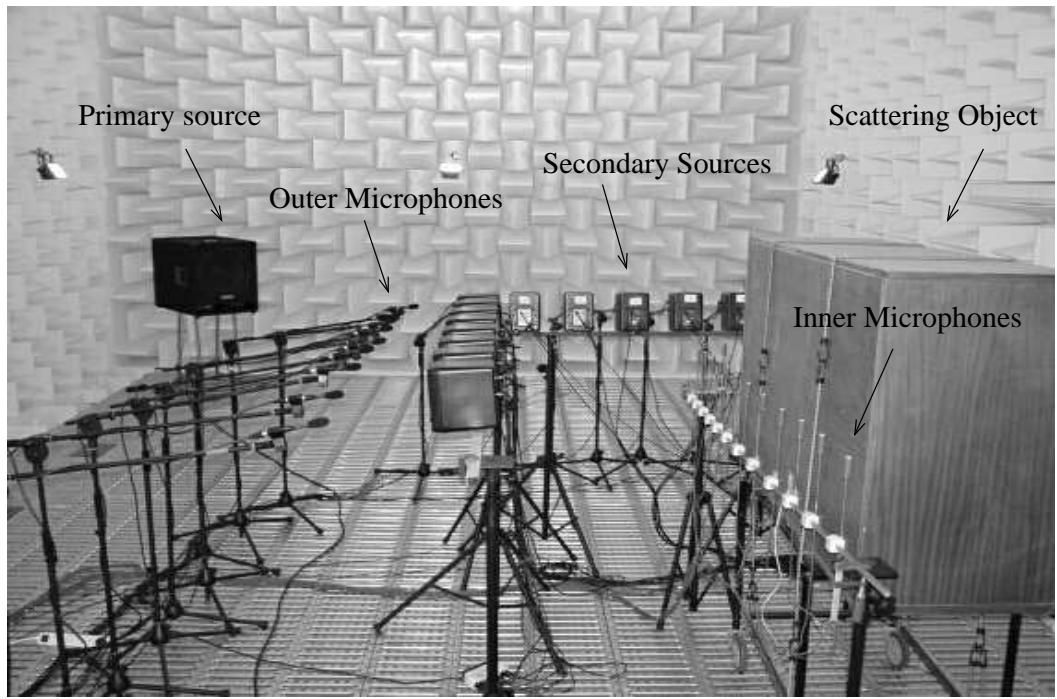}  
\vspace*{2cm} \caption{An experimental set-up for active control of the scattered noise}  
\label{photo}
\end{center}  
\end{figure}

The laboratory experiment of active control of the scattered noise
was carried out in the LMA large anechoic chamber. The
scattering object was a parallelepipedic wooden box with dimensions 
$1.56\times 0.94\times 0.5\;\textrm{m}^3$, the same as for the rigid object considered
for numerical simulations in section~3.
The incident noise was a 280~Hz pure-tone generated by a  primary loudspeaker.
For real-time active control, 18 inner microphones and 14 loudspeakers 
were used to minimize the scattered field on 16 outer microphones.
This experimental set-up is shown in Fig.~\ref{photo} and 
the distances between actuators and sensors are reproduced in Fig.~\ref{schemamanip}.
A metallic floor was left in the chamber for the experiments in order to remove or insert
the scattering object without difficulty. This meant that the chamber was not truly
anechoic but, when the control strategy was implemented, free-field conditions were not
required because the filters accounting for the scattering
of the body were identified off-line, as explained in section~2. However
a quantitative comparison of the experimental results to the 
free-field simulations in section~3 was not possible.
Numerical implementations only provided an idea of the expected 
performances of the real-time active control experiment.

The experimental set-up shown in Fig.~\ref{photo} does not pretend to be a reduced-scale
of a practical implementation such as a submarine surrounded by anti-noise sources.
The size and the shape of a real submarine, as well as the frequency range of sonars,
make active control of echoes from submarines an extremely difficult task.
Furthermore, the size and the position of the control
sources in the experiment make them significantly scatter the primary noise, whereas
the control strategy under testing aims at reducing the noise scattered by the parallelepiped only;
this scattering by the secondary source could not be accepted in the case of a submarine.
In an other application such as active control of wall reflections in anechoic chambers,
the secondary sources could be embedded in the walls, which would prevent them from increasing
the scattering by the walls. However that may be, the set-up shown in Fig.~\ref{photo} was designed
to implement and to test the control strategy in real-time, regardless of a specific application;
the aim of the experiment was merely to demonstrate that real-time active control of scattered radiation
could be performed in 3D.

\subsection{Preliminary measurements and calculations}

As indicated in section~2, real-time active control of the scattered noise
requires preliminary measurements in order to identify the scattering filters
and the secondary paths.
In practice this amounts to measuring the acoustic pressure at all the sensors 
(inner and outer microphones) for all the secondary sources successively.
This pressure is measured with and without the scattering object and
the scattered pressure at the outer microphones is obtained by subtracting
the pressure measured without the object to that measured with the object.

\begin{figure}[!ht] 
\begin{center} 
\includegraphics[width=14cm]{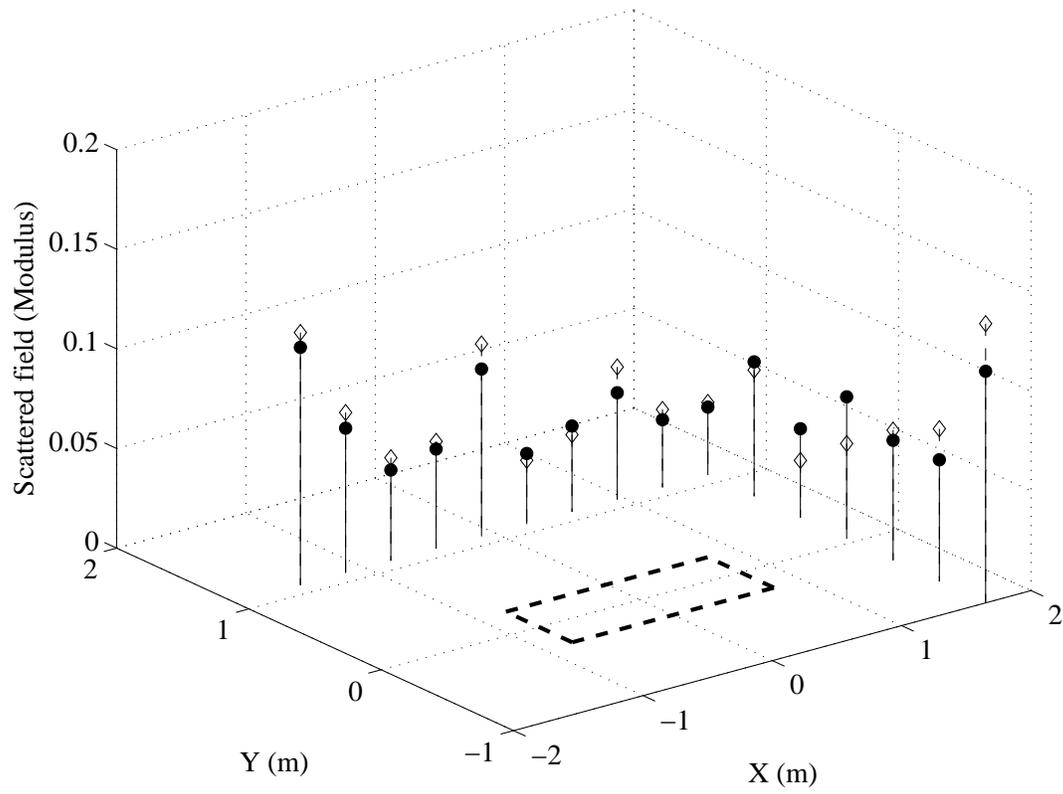}    
\vspace*{2cm} \caption{Magnitude of the measured (--- $\bullet$) and estimated
 (- - $\diamond$) scattered noise as a function of microphone position}  
\label{estime2_3d}  
\end{center}  
\end{figure}

\begin{figure}[!ht] 
\begin{center} 
\includegraphics[width=14cm]{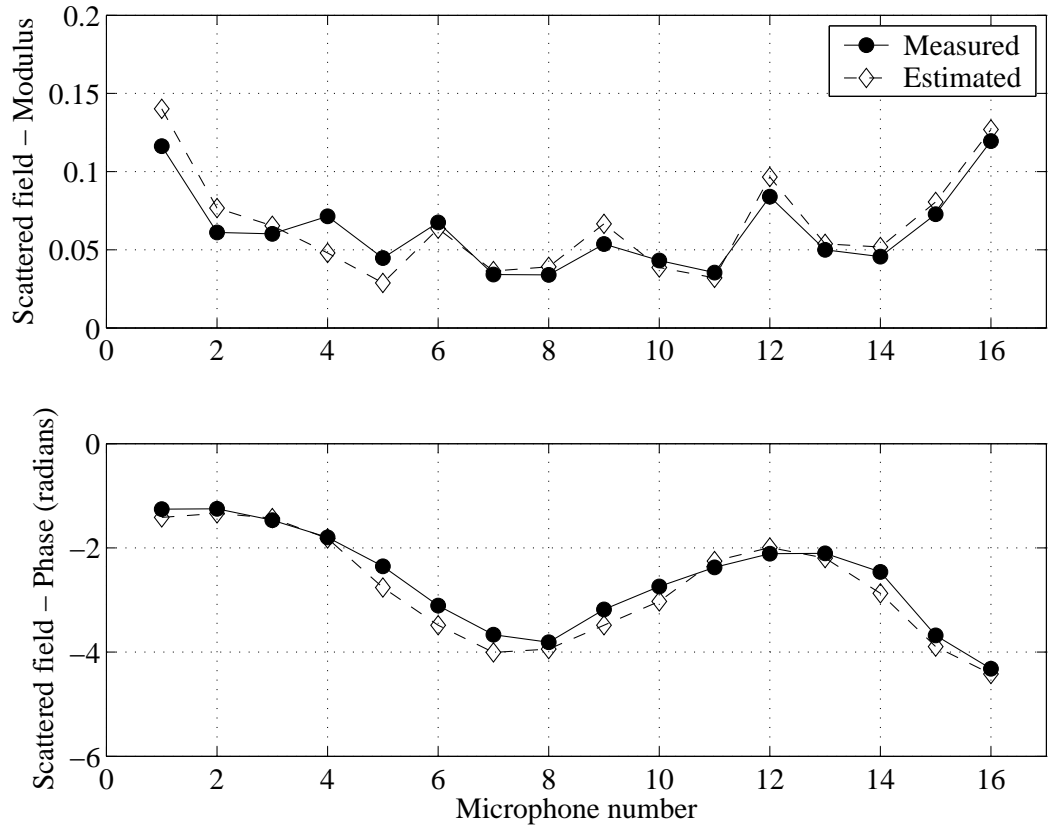}    
\vspace*{2cm} \caption{Magnitude and Phase of measured (--- $\bullet$) and estimated
 (- - $\diamond$) scattered noise}   
\label{estime2}  
\end{center}  
\end{figure}

At the first attempt, the filters $\hat{g}_k$, accounting for scattering,
were obtained by direct inversion of Eq.~(\ref{eq:Green3}) evaluated at the
outer microphones for each of the secondary loudspeakers.
With these coefficients, the scattered noise from the primary source,
estimated from the total noise, was compared to the measured scattered noise.
It appeared that in this way the primary scattered noise was estimated with
a 130\% error, which was unacceptable for active control. In fact, the inversion
of Eq.~(\ref{eq:Green3}) appeared to be ill-conditioned and required
some precaution. Singular Value Decomposition associated to Tikhonov regularisation
was used to extract a regularized pseudo-inverse and the regularized scattering operator
produced an estimate of the scattered pressure field with an acceptable error of at most 30\%
(see Fig.~\ref{estime2_3d} and Fig.~\ref{estime2}).

\subsection{Real-time control results}

The real-time active control strategy was implemented on the LMA multiprocessor
system COMPARS. The sampling frequency selected for the experiments at 280~Hz was 
1100~Hz. In principle, control of pure-tone noise requires FIR filters
with only 2 coefficients for modelling the secondary path and the scattering
filters. In practice however, using a few more coefficients sometimes improves the
convergence of the adaptive control algorithm and the experiments were more-or-less casually 
conducted with 6-coefficient FIR filters.

\begin{figure}[!ht] 
\begin{center}  
\includegraphics[width=14cm]{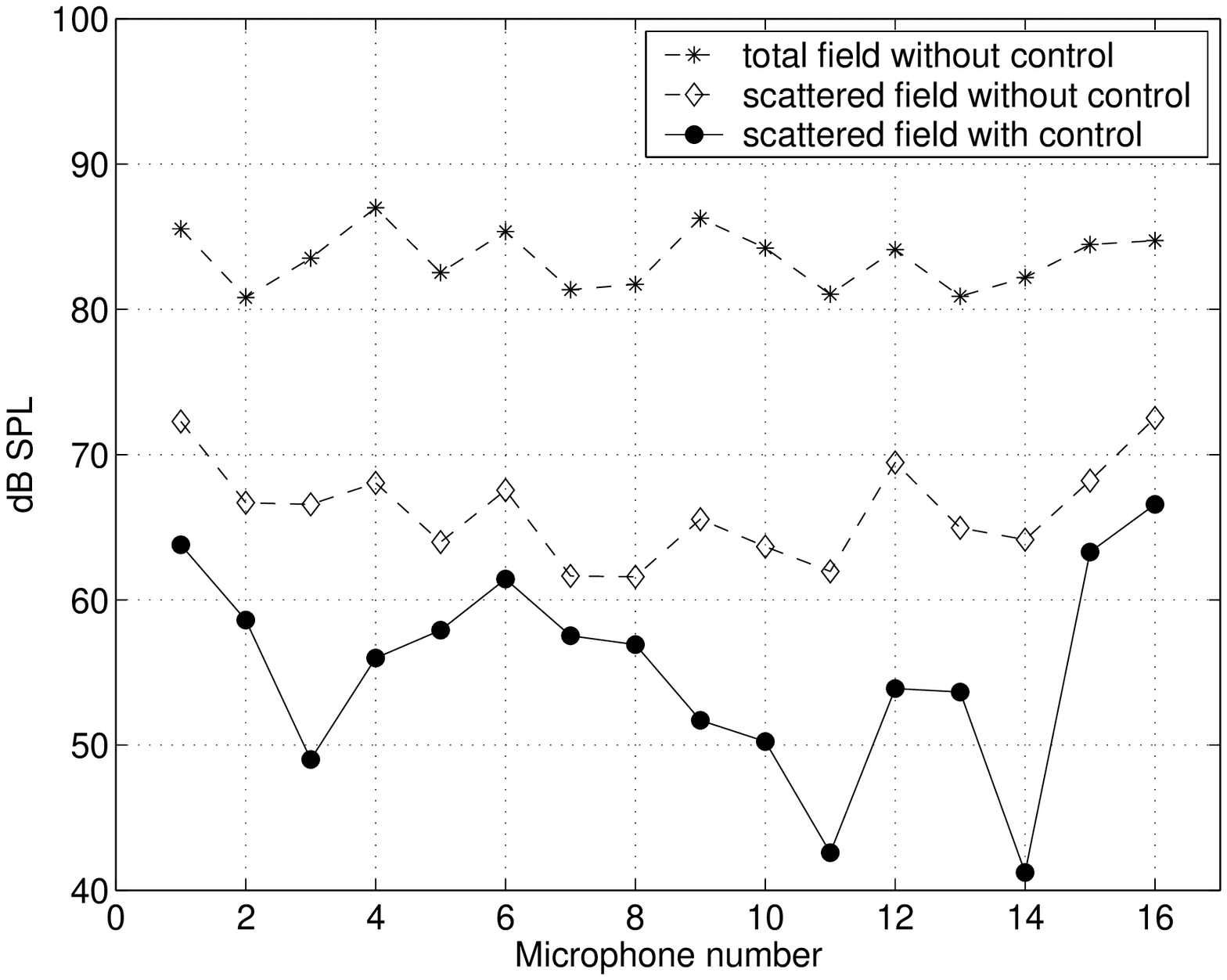} 
\vspace*{2cm} \caption{Magnitude of the total (- - $*$) and  scattered noise 
(- - $\diamond$) without control and modulus of the scattered noise with control (--- $\bullet$)}  
\label{result1}  
\end{center} 
\end{figure}

The electric signal fed to the primary source was used as a reference signal
for the FELMS algorithm. Such a signal ensures control stability provided the FELMS
algorithm does not diverge. In the case of a pure-tone excitation, an acoustic reference signal
may be selected because the primary noise is fully predictible, but this entails
the risk of control unstability because of acoustic feedback. Since
the experiments shown in this paper were focusing on scattered radiation control,
a safe electric reference signal was adequate. However, Fig.~\ref{result1} will show
that the total noise measured by the minimization microphones was only very slightly
affected by control of the scattered noise. It retrospectively appears that
an acoustic reference signal could have been selected for feedforward control
because the level of acoustic feedback during control would have been very low.

The FELMS adaptive algorithm can be applied to the control of broadband noise.
However, in this case, the causality constraint may drastically restrains the
performances or active noise control even if an electric reference signal is
available (see \emph{e.g.} \cite{Friot-CFA}). The convergence of the FELMS
may also be problematic for random noises with low dips in the power spectral density (see \cite{Elliott}). 
Since the purpose of this study was merely to demonstrate scattered noise control,
broadband implementation, with its specific difficulties, was discarded from the beginning
of the study.
However, it must be noted that the increased computational burden, due to the
longer FIR filters required in the case of broadband noise, was not the reason
why broadband implementation was rejected, because the strategy implemented in this
paper requires only twice as much processing time and memory as a standard FELMS algorithm.

\begin{figure}[!ht] 
\begin{center}
\includegraphics[width=14cm]{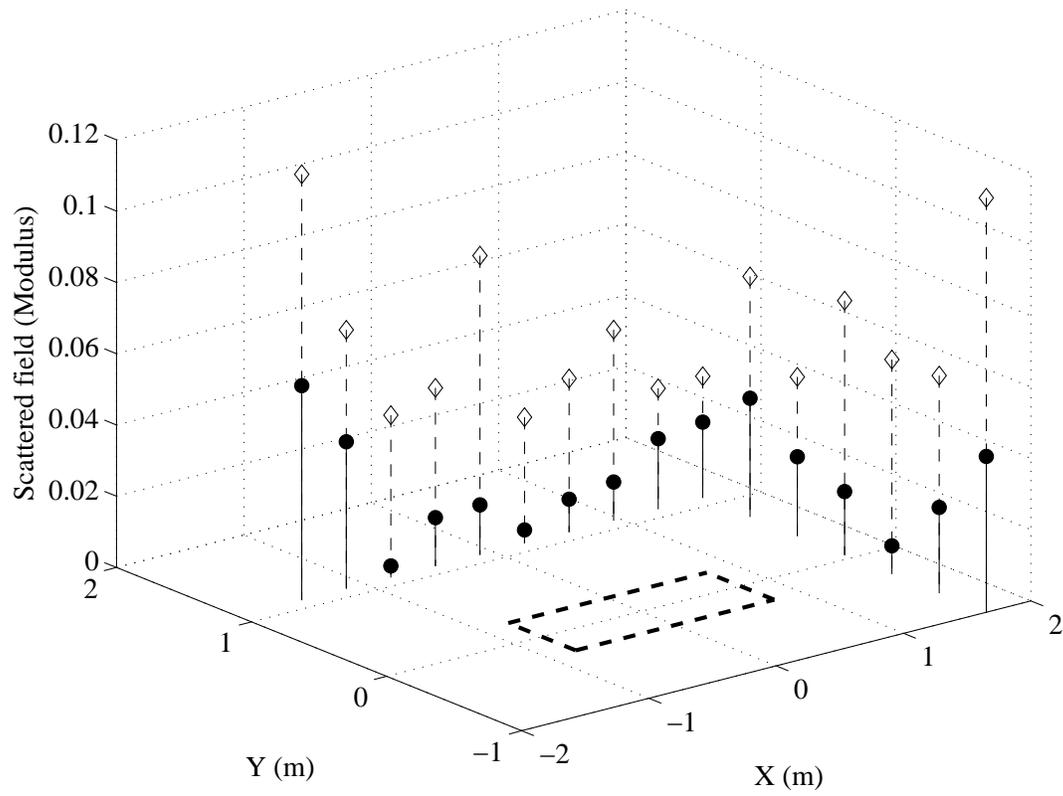}
\vspace*{2cm} \caption{Magnitude of the measured scattered noise with (--- $\bullet$)
and without (- - $\diamond$) control}
\label{result1_3d}  
\end{center} 
\end{figure}

The total pressure was measured at 280~Hz with and without control
at the 16 outer microphones. The scattered pressure was obtained
by substracting the pressure without the object to the pressure with the object.
The scattered pressure with and without control is shown
in Fig.~\ref{result1_3d} and in Fig.~\ref{result1}. Control reduced
the scattered noise by at least 5~dB at all the error microphones. The
attenuation was larger than 20~dB at a few error microphones. The level of
the scattered field without control was large at these latter microphones,
which probably led to a better estimation of the error signal;
least-square minimization also preferably minimizes the largest error signals.

Fig.~\ref{result1} also displays the total noise without control
at the error microphones. The scattered noise was only a fraction of the total noise,
which makes identification and control of the scattered noise difficult in practice and
may explain the 30\% error on the estimated error signal. It can also be noted
that perfect cancellation of the scattered noise would not significantly change the total noise level.  
However, the noise reduction achieved in practice is of the same magnitude
as the reduction predicted by the numerical simulations in section~3. In the end
the authors considered that the achieved control results were reasonably good for the given
arrangement of sensors and actuators.

\section{Conclusion}

This paper has shown experimental results of active control of the noise scattered in 3D
by a parallelepiped using ordinary sources, microphones and digital signal processing units.
Although in the experiment the scattered noise had a level of only one
tenth of the total noise level at the error sensors, a 10~dB reduction of the
scattered noise at the sensors was achieved, which was close to the optimal control
for the given arrangement of sources according to the numerical simulations.
The control strategy required no information about the incident wave.
All the Frequency Response Functions required for real-time computations were identified before
control from noise measurements with the secondary sources only.

Therefore the control strategy introduced in \cite{Friot} was successfully implemented in
a 3D case. However the applicability of this strategy to problems of industrial
interests remains to be demonstrated. Application to submarines is difficult,
not to say unrealistic, because of the required number of channels.
The application currently being investigated by the authors is the attenuation
of the reflections on the walls of an anechoic chamber, which is difficult to achieve
at very low frequency with passive absorption. In this case active control
of the scattered noise require a tractable number of actuators and sensors.
However the filters mapping the total noise to the scattered noise could not be identified
from preliminary measurements involving the removal of the reflecting surface, as it
was the case in this paper. Instead, temporal windowing could be used in order
to separate the incident noise from the scattered noise, but this would require
a perfectly equalized and calibrated low-frequency source in order
to separate the wall reflections from the echoes due to the source dynamics.

\newpage 

\begin{table} [!h]
{\small 
\begin{center}
TABLE I: Parameters of various studied configurations\\
\ \\
\begin{tabular}{ccccccc}
& Number of  & Number of &  Number of & estimation error & control error & Figures\\
& inner sensors & outer sensors & actuators &
$\| \sum_{k} g_k p(\z{r}_k) - \z{p}_{s} \| $ 
& $\| \z{e}\|$ & \\
\hline \hline\\
conf.(1) & 54 & 324 & 302 & 2.4 \% & 5.4 $10^{-9}$ \% & \ref{2803d}
\\
conf.(2) & 54 & 58 & 56 & 5.4 \% & 1.6 \% & \ref{280deg1} \\
conf.(3) & 18 & 16 & 14 & 19 \% & 0.44 \% & \ref{280deg2}
\end{tabular}
\end{center}}
\end{table}
\newpage 
 
\newpage
 
\newpage 

\newpage 

\newpage 

\newpage 
 
\newpage 
 
\newpage 
 
\newpage 

\newpage 

\newpage 

\newpage


\begin{thebibliography}{20}

\bibitem{Nelson}
Nelson,~P.~A., and Elliott,~S.~J.~({\bf 1992}).
{\it Active Control of Sound} (Academic Press, London), pp.~290--293.

\bibitem{Mangiante}
Mangiante~G., and Roure~A.~({\bf 1994}).
"Autodirective sources for 3D Active Noise Control,"
{\it Proc. Internoise 94}, pp.~{1293--1298}.

\bibitem{Howarth}
Howarth~T.~H., Varadan~V.~K., Bao~X., and Varadan~V.~V.~({\bf 1992}).
"Piezocomposite coating for active underwater sound reduction,"
{\it J. Acoust. Soc. Am.}  {\bf 91}, pp.~{823--831}.

\bibitem{Nayrolles}
Nayrolles~B., and Nicolas~T.~({\bf 1997}).
"Conception of an active anechoic multicellular layer," 
{\it Proc. Active 97}, pp.~{1179--1188}.

\bibitem{Olson}
Olson~H.~F., and May E. G. ({\bf 1953}).
"Electronic sound absorber,"
{\it J. Acoust. Soc. Am.}  {\bf 25}, pp.~{1130--1136}.

\bibitem{Orduna}
Orduna-Bustamante~~F., and Nelson P. A. ({\bf 1992}).
"An adaptive controller for the active absorption of sound,"
{\it J. Acoust. Soc. Am.}  {\bf 91}, pp.~{2740--2747}.

\bibitem{Furstoss}
Furstoss~M., Thenail~D., and Galland~M.-A.~({\bf 1997}).
"Surface impedance control for sound absorption: direct and hybrid
passive/active methods,"
{\it J. Sound Vib.} {\bf 203}, pp.~{219--236}. 

\bibitem{Scandrett}
Scandrett C.~L., Shin Y.~S., Hung K.~C., Khan M.~S. and Lilian C.~C~({\bf 2004}).
"Cancellation techniques in underwater scattering of acoustic signals,"
{\it J. Sound Vib.} {\bf 272}, pp.~{513--537} 

\bibitem{Friot}
Friot~E., and Bordier~C.~({\bf 2004}).
"Real-time active suppression of scattered acoustic radiation,"
{\it J. Sound Vib.}, {\bf 278(3)}, pp.~563--580

\bibitem{Friot-CFM}
Friot E., and Guillermin~R.~({\bf 2005}).
"Assourdissement actif \`a basse-fr\'equence des chambres an\'echo\"{\i}ques,"
{\it Proc. CFM 2005}, {on CD-ROM}.

\bibitem{Elliott}
Elliott~S.~J.~({\bf 2001}).
{\it Signal Processing for Active Control} (Academic Press, London), pp.~142-144 and pp.~251-254.

\bibitem{Morse}
Morse~P.~M., and Ingard~T. K. U. ({\bf 1968}).
{\it Theoretical Acoustics} (Princeton University Press, Princeton), p.~321.

\bibitem{Ise}
Ise S. ({\bf 1999}).
"A principle of sound field control based on the Kirchoff-Helmholtz integral
equation and the theory of inverse systems,"
{\it Acta Acust.} {\bf 85}, pp.~78--87.  

\bibitem{Kirkup}
Kirkup S. ({\bf 1998}).
{\it The Boundary Element Method in Acoustics} (Integrated Sound Software, Hebden Bridge).

\bibitem{Tikhonov}
Tikhonov A.~N., and Arsenin V.~Y. ({\bf 1977}).
{\it Solutions of Ill-Posed Problems} (Winston \& Sons, Washington D.C.).

\bibitem{Friot-CFA}
Friot~E.~({\bf 2004})
"Causality Constraints in Multi-Channel Active Control of Random Noise,"
{\it Proc. CFA/DAGA'04}, on CD-ROM.

\end{thebibliography}
\end{document}